\documentclass[a4paper,11pt]{article}
\usepackage{jheppub} 
\usepackage{lineno}
\usepackage{subcaption}
\usepackage{xcolor}

\title{Soliton-Black Hole Phase Transitions in the Presence of String Clouds}

\author[a,b,c]{Ziqing Chen}
\author[c,d,e]{Robert B. Mann}
\affiliation[a]{Lanzhou Center for Theoretical Physics, Key Laboratory of Theoretical Physics of Gansu Province, School of Physical Science and Technology, Lanzhou University, Lanzhou 730000, China}
\affiliation[b]{Institute of Theoretical Physics and Research Center of Gravitation, 	Lanzhou University, Lanzhou 730000, People's Republic of China}
\affiliation[c]{Department of Physics and Astronomy, University of Waterloo, Waterloo, Ontario, N2L 3G1, Canada}
\affiliation[d]{Institute for Quantum Computing, University of Waterloo, Waterloo, Ontario, N2L 3G1, Canada}
\affiliation[e]{Perimeter Institute for Theoretical Physics,  Waterloo, Ontario, N2L 2Y5, Canada}

\emailAdd{rbmann@uwaterloo.ca}

\abstract{We study soliton-black hole phase transitions in asymptotically AdS planar spacetimes sourced by a cloud of strings. In the planar background, the string cloud exhibits a brush-like configuration, where the strings are aligned parallel to each other and extend along the radial direction. This leads to a stress tensor with nonvanishing components only along the temporal and radial directions. By comparing the Euclidean on-shell actions of the planar black hole and the corresponding AdS soliton under the same boundary conditions, we obtain the free energy difference between the two phases. Our results show that the string cloud parameter significantly modifies the competition between the   black hole phase and the   soliton phase. In particular, positive and negative values of the string cloud parameter affect the phase structure in different ways, changing the dominance relation between the black hole and soliton configurations.}

\begin{document}
\maketitle
\flushbottom

\section{Introduction}
\label{sec:intro}
Gauge/gravity duality provides a powerful framework for studying strongly coupled systems through classical gravitational dynamics in asymptotically anti-de Sitter (AdS) spacetimes ~\cite{witten1998antisitterspaceholography,Maldacena_1999,Gubser_1998}. One of most important applications is black hole phase transitions,  where the cosmological constant is treated as the pressure~\cite{Bekenstein1973,kastor2009enthalpy}. The Hawking-Page transition is one of the earliest and most important examples of black hole phase transitions in asymptotically AdS spacetimes~\cite{hawking1983thermodynamics}. It describes a transition between thermal AdS space and an AdS black hole and is dual to the confinement/deconfinement transition in the boundary gauge theory~\cite{witten1998antisitterspaceholography}. Motivated by this discovery, many new thermodynamic phase structures have been identified, including small/large black hole phase transitions, reentrant phase transitions, and a variety of critical phenomena in AdS spacetimes~\cite{altamirano2013reentrant,altamirano2014kerr,dolan2014isolated,frassino2014multiple,cai2013pv,hennigar2017superfluid,kubizvnak2017black,zou2017reentrant,tavakoli2022multi}.

Closely related to the Hawking--Page transition is the phase transition between the AdS soliton and the planar AdS black hole. Originally proposed by Horowitz and Myers~\cite{Horowitz_1998}, the AdS soliton is obtained through a double Wick rotation of the planar AdS black hole and is interpreted holographically as a confining vacuum of the dual gauge theory. In contrast, the planar AdS black hole corresponds to a deconfined thermal phase. It was subsequently shown that these two geometries undergo a first-order phase transition when their boundary periodicities are appropriately matched~\cite{surya2001phase}. Since then, the soliton--black hole transition has been extensively investigated in a variety of gravitational theories, revealing a rich thermodynamic phase structure and providing valuable insights into confinement/deconfinement phenomena~\cite{cai2007ricci,eune2013hawking,banerjee2007phase,copsey2006gravity,anabalon2022phase,quijada2024triple,chen2008thermodynamics,Ahmed:2026fes}.

String clouds offer a simple phenomenological way to incorporate a distribution of fundamental strings into the gravitational background. A cloud of strings is an effective matter model in which many one-dimensional strings are treated as a continuous source of gravity. Unlike ordinary dust made of point particles, a string cloud has both energy density and tension along the direction of the strings. This model was first developed in the context of general relativity by Stachel and Letelier, and Letelier later used it to study cosmological models with massive strings~\cite{letelier1979clouds,stachel1980thickening,letelier1983string}. Since then, string clouds have been applied in many gravitational systems, especially in black-hole physics and modified gravity, where they are used to study how string-like matter can change the spacetime geometry, thermodynamics, geodesic motion, shadows and quasinormal modes\cite{herscovich2010black,Ghosh_2014,lee2015lovelockblackholethermodynamics,Li_2021,Chabab_2020,C_rdenas_2021,Gogoi_2022}.

The backreaction of the string cloud modifies the bulk metric and therefore changes the thermodynamic behavior of black holes and soliton geometries. In particular, the string cloud parameter contributes directly to the regulated on-shell action. Therefore the phase structure is controlled not only by the horizon radius or the soliton edge, but also by the relative strength of the string cloud in the competing geometries. If the black hole and the soliton carry different string cloud parameters, the string cloud contribution contains a term linear in the cutoff radius. In the large-$R$ cutoff limit this term diverges, causing the free energy difference to become ill-defined. Only when the string cloud parameters are identified does the divergent contribution cancel, leading to a finite free energy difference and a consistent phase structure.

Here we investigate how the presence of a string cloud affects the soliton-black hole phase structure. We compare the relevant gravitational saddles under the same boundary conditions and determine the transition point from their renormalized Euclidean actions. This provides a controlled way to study how string cloud influence the confinement and deconfinement phases encoded by the soliton and black hole geometries. We also study the effects of negative string cloud parameters on the black hole-soliton phase transition. This allows us to explore a wider class of thermodynamic behaviors and examine how the sign of the string cloud contribution modifies the competition between the two phases.

The remainder of our  paper is organized as follows. In Sec. II, we introduce the general setup of the system, including the gravitational action, the string cloud configuration, and the corresponding planar black hole and AdS soliton solutions. In Sec. III, we compute the renormalized Euclidean actions and obtain the free energy difference between the two geometries. We then investigate the effects of both positive and negative string cloud parameters on the phase structure in Sec. IV. Finally, we summarize our results and discuss possible future directions in Sec. V.

\section{Setup and Background Solutions}

In this section, we briefly introduce the setup, including the action for the background with string clouds, the configuration of the string cloud in planar spacetime, and the corresponding metric.

We consider Einstein gravity with a negative cosmological constant and string clouds. The total Euclidean action may be written as

\begin{equation}\label{eq:action_simplification}
  I = I_{\mathrm{bulk}} + I_{\mathrm{GHY}}  + I_{\mathrm{s}},
\end{equation}
where  $I_{\mathrm{GHY}}$ is the is bulck term, $I_{\mathrm{GHY}}$ is the Gibbons-Hawking-York boundary term, and $I_{\mathrm{s}}$ is the string clouds action. The bulk term is
\begin{equation}\label{eq:action_bulk}
  I_{\mathrm{bulk}}
  = -\frac{1}{16\pi G}\int d^4x \sqrt{g}\left(R+\frac{6}{L^2}\right)
\end{equation}
where  $L$ is the AdS radius. 

As for the string cloud term, it is contributed by several individual strings, which can be described by the Nambu--Goto action~\cite{letelier1979clouds}:
\begin{equation}\label{eq:action_NG}
  I_{\mathrm{NG}}
  = -T_{\mathrm{s}}\int d^2\xi \sqrt{-\gamma},
  \qquad
  \gamma_{ab}=g_{\mu\nu}\frac{\partial x^\mu}{\partial \xi^a}
  \frac{\partial x^\nu}{\partial \xi^b}.
\end{equation}
It is often useful to introduce the bivector
\begin{equation}\label{eq:bivector}
  \Sigma^{\mu\nu}
  = \epsilon^{ab}\frac{\partial x^\mu}{\partial \xi^a}
  \frac{\partial x^\nu}{\partial \xi^b},
  \qquad
  \sqrt{-\gamma}
  = \sqrt{-\frac{1}{2}\Sigma^{\mu\nu}\Sigma_{\mu\nu}}.
\end{equation}
The total contribution of these individual string actions can be expressed as the string cloud action
\begin{equation}\label{eq:stringact}
  I_{\mathrm{s}}
  = -\int d^4x\sqrt{-g}\rho(x)
  \sqrt{-\frac{1}{2}\Sigma^{\mu\nu}\Sigma_{\mu\nu}},
\end{equation}
where $\rho(x)$ is the proper density of strings. 

Varying the string action \eqref{eq:stringact}
with respect to the metric in general yields an anisotropic stress tensor.  However for a static cloud of   strings whose density is isotropic 
we obtain
\begin{equation}\label{EMtensor}
  T^\mu{}_\nu = \rho(r)\,\mathrm{diag}(-1,-1,0,0).
\end{equation}
We emphasize that the discussion above does not rely on assuming either spherical or planar symmetry of the spacetime. Therefore, even in the planar case, the form of the stress tensor remains identical to that in the spherically symmetric background. Nevertheless, although the stress tensor takes the same form, the geometric configuration of the string cloud changes significantly in the planar spacetime. Instead of the hedgehog-like radial distribution appearing in the spherical case \cite{Headrick:2007ca}, the strings in the planar geometry exhibit a brush-like configuration, aligned parallel to each other along the radial direction.

\begin{figure}[htbp]
    \centering

    \begin{subfigure}{0.45\textwidth}
        \centering
        \includegraphics[width=\textwidth]{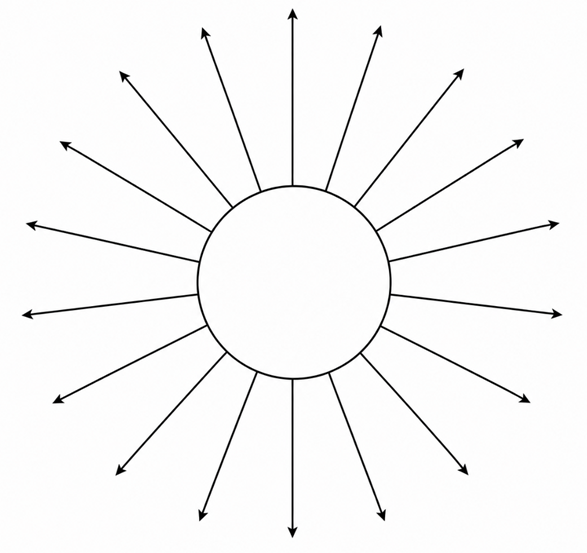}
        \caption{Spherical case}
    \end{subfigure}
    \hfill
    \begin{subfigure}{0.45\textwidth}
        \centering
        \includegraphics[width=\textwidth]{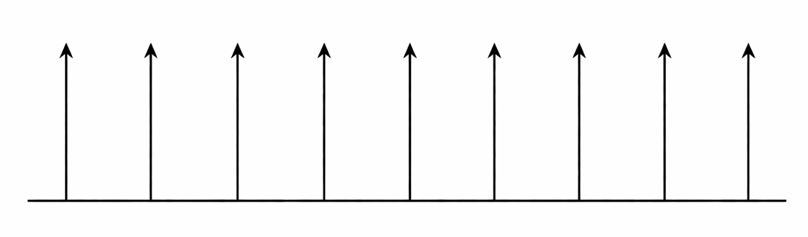}
        \caption{Planar case}
    \end{subfigure}

    \caption{Comparison between the spherical and planar cases. In Fig.~(a), corresponding to the spherical case, the strings extend radially outward from the horizon toward infinity, forming a hedgehog-like configuration. In Fig.~(b), corresponding to the planar case, the strings emerge from the horizon and remain parallel to each other, forming a brush-like configuration.}
    \label{fig:1}
\end{figure}

The conservation equation $\nabla_\mu T^\mu{}_\nu=0$ fixes the radial density profile. In four-dimensional planar coordinates, this gives

\begin{equation} \label{eq:stringdensity}
  \rho(r)=\frac{b}{r^2},
\end{equation}
where $b$ is the string cloud parameter. This parameter measures the strength of the cloud. Surprisingly, the resulting expression coincides with that obtained in the spherically symmetric case. 

The planar black hole ansatz may be written as
\begin{equation}\label{eq:BH_line_element}
  ds_{\mathrm{BH}}^2
  = -f_{\mathrm{b}}(r)dt^2+\frac{dr^2}{f_{\mathrm{b}}(r)}
  +r^2(dx^2+dy^2),
\end{equation}
where the horizon is located at $r=r_{\mathrm{b}+}$ with $f_{\mathrm{b}}(r_{\mathrm{b}+})=0$. The horizon is a constant-$r$ surface with flat planar geometry in the $(x,y)$ directions. The regularity of the Euclidean black hole fixes the periodicity of $t$, which is given by $\beta_b$, while $\eta_{b}$ is (arbitrary) period of  $x$. 

The AdS soliton geometry is obtained by a double Wick rotation, which exchanges the role of the Euclidean time circle and one compact spatial circle. The line element of AdS soliton solution is
\begin{equation}\label{eq:SOL_line_element}
  ds_{\mathrm{sol}}^2
  = -r^2dt^2+\frac{dr^2}{f_{\mathrm{s}}(r)}
  +f_{\mathrm{s}}(r)dx^2+r^2dy^2,
\end{equation}
where the spacetime smoothly caps off at $r=r_{\mathrm{s}+}$ with $f_{\mathrm{s}}(r_{\mathrm{s}+})=0$. The regularity at the soliton edge (or tip) fixes the periodicity of $x$, which is given by $\beta_s$.

Solving the Einstein equations in each case, the corresponding black hole and soliton metrics are
\begin{equation}\label{eq:metrics}
\begin{aligned}
f_b(r) &= \frac{r^2}{\ell^2}-b_b-\frac{2M_b}{r}, \\
f_s(r) &= \frac{r^2}{\ell^2}-b_s-\frac{2M_s}{r}.
\end{aligned}
\end{equation}
The periods $\beta_b$ and $\beta_s$ are obtained from the regularity conditions and can be expressed in terms of the first derivatives of the corresponding metric functions
\begin{equation}\label{eq:inverse_T}
\begin{aligned}
\beta_{b}
=\frac{4\pi}{f'_{b}(r_{b+})}
=\frac{4\pi r_{b+}\ell^2}{3r_{b+}^{2}-b_{b}\ell^{2}} .\\
\beta_{s}
=\frac{4\pi}{f'_{s}(r_{s+})}
=\frac{4\pi r_{s+}\ell^2}{3r_{s+}^{2}-b_{s}\ell^{2}} .
\end{aligned}
\end{equation}
For the black hole this is the inverse temperature, and for the soliton the analogous expression fixes the period of the circle that shrinks at the tip. 

\section{On-shell Action and Free Energy}

For the same asymptotic boundary conditions, the planar AdS black hole and the AdS soliton can be regarded as two competing configurations within the same thermal ensemble. Under fixed boundary conditions, both solutions satisfy the Einstein field equations and may exist as equilibrium states. However, the preferred phase is determined by its free energy, with the configuration of lower free energy being thermodynamically dominant~\cite{surya2001phase}.

In the Euclidean path integral formalism, the free energy is directly related to the Euclidean on-shell action. Therefore, to determine the dominant phase, we calculate the on-shell actions of both the black hole and soliton solutions and compare their corresponding free energies. The sign of the free energy difference determines the thermodynamically preferred phase. If the black hole has a lower free energy, the black hole phase dominates; conversely, if the soliton has a lower free energy, the soliton phase dominates. The point at which the free energy difference changes sign corresponds to the phase transition point.

Motivated by this, in this section we compute the Euclidean on-shell actions and free energies of the planar AdS black hole and AdS soliton, and investigate how the presence of a string cloud modifies the phase structure of the system.

Before comparing the Euclidean on-shell actions of the black hole and AdS soliton solutions, we must ensure that the two configurations share the same boundary conditions. Otherwise, if they belong to different thermal ensembles, their free energies cannot be meaningfully compared. Therefore, we employ the standard matching procedure, in which the induced metrics of the two geometries are required to coincide at a sufficiently large cutoff radius $r=R$. Specifically, the metric components along the Euclidean time and spatial $x$ directions are matched, leading to relations between the corresponding periods of the black hole and soliton solutions. The matching condition is:
\begin{equation}\label{eq:inverseT_matching}
\beta_b \sqrt{f_b(R)} = R \eta_s,
\qquad
\beta_s \sqrt{f_s(R)} = R \eta_b.
\end{equation}
where $\eta_{s}$ is the period of the
Euclidean time $\tau=-it$ for the soliton
\eqref{eq:SOL_line_element}.
In the limit $R\to\infty$, we obtain
\begin{equation} \label{eq:periodicity_matching}
  \eta_{\mathrm{s}}=\frac{\beta_{\mathrm{b}}}{\ell},
  \qquad
  \eta_{\mathrm{b}}=\frac{\beta_{\mathrm{s}}}{\ell},
\end{equation}
It is straightforward to verify that  \eqref{eq:periodicity_matching} can still be satisfied in the limit $R\to\infty$, even when $M_b \neq M_s$ and $ b_b \neq b_s$.

The Euclidean action for Einstein gravity coupled to a string cloud is given by,
\begin{align}
I
=
&
-\frac{1}{16\pi G}
\int_{\mathcal M} d^{4}x \sqrt{g}\,(R-2\Lambda)
-\frac{1}{8\pi G}
\int_{\partial\mathcal M} d^{3}x \sqrt{h}\,K\notag
\\
&
-\int d^4x\sqrt{-g}\rho(x)
 \sqrt{-\gamma},.
\end{align}
For convenience, we set $8\pi G = 1$ in the following calculations.

The dominant phase is determined by the sign of the free energy difference between the two saddle points. Equivalently, one may compare the renormalized Euclidean actions~\cite{surya2001phase},

\begin{equation}\label{eq:Delta_I}
\begin{aligned}
      \Delta I
  &
  = I_{\mathrm{BH}}-I_{\mathrm{SOL}}
  \\
  &
  =\frac{3\,\mathrm{VOL}(R/\Gamma)}{l^2}\left(\beta_b\eta_b\int_{r_{b+}}^{R} r^{2}\,dr-\beta_s\eta_s\int_{r_{s+}}^{R} r^{2}\,dr\right)
  \\
  &
  +\mathrm{VOL}(R/\Gamma)\left(\beta_s\eta_s\int_{r_{s+}}^{R} b_{s}\,dr-\beta_b\eta_b\int_{r_{b+}}^{R} b_{b}\,dr\right),
\end{aligned}
\end{equation}
where  $\mathrm{VOL}(R/\Gamma)$ denotes the volume obtained by integrating over the transverse $y$-direction, with $\Gamma$ denoting periodic identification in the $y$
direction. In the large-$R$ limit, the 
 divergent part of 
  the second line in \eqref{eq:Delta_I}
  (the Gibbons-Hawking term) 
vanishes 
due to \eqref{eq:periodicity_matching}.

 For clarity, we emphasize the meaning of the various periodicities. The quantity $\beta_b$ denotes the period of the Euclidean time coordinate $t$ in the black hole geometry, while $\eta_b$ denotes the period of the spatial coordinate $x$ in the black hole geometry. For the AdS soliton, $\beta_s$ denotes the period of the spatial coordinate $x$, whereas $\eta_s$ denotes the period of the Euclidean time coordinate $t$.

It is convenient to  work with the action density and free energy density, obtained by dividing the overall volume factor $\mathrm{VOL}(R/\Gamma)$
\begin{equation}
\Delta \mathcal I = \frac{\Delta I}{\mathrm{VOL}(R/\Gamma)},
\end{equation}
 since this quantity is finite even if the transverse $y$ direction is not periodically identified. 
The free energy density is
\begin{equation} \label{eq:delta_F}
 \Delta F = \frac{\Delta \mathcal I}{\beta_b},
\end{equation}
and 
the black hole phase dominates when $\Delta F<0$. Otherwise   the soliton phase dominates when $\Delta F>0$; if $\Delta F=0$ the two phases are in equilibrium.

In the absence of a string cloud, the transition is controlled by the relative size of the black hole horizon and the soliton tip after the boundary periodicities are matched. In the reference subtraction method, the pure gravitational contribution is proportional to $r_{\mathrm{s}+}^{3}-r_{\mathrm{b}+}^{3}$,  reproducing the usual result that the sign changes when the black hole and soliton scales are interchanged.

With the string cloud included, the regulated free energy density obtained by substituting the above relations is
\begin{equation}
  \begin{aligned}
  \Delta F(R)
  =
  \frac{\beta_{\mathrm{s}}}{\ell^3}
  \left(r_{\mathrm{s}+}^{3}-r_{\mathrm{b}+}^{3}\right) +\frac{\beta_{\mathrm{s}}}{\ell}
  \left(b_{\mathrm{b}}-b_{\mathrm{s}}\right)R +\frac{\beta_{\mathrm{s}}}{\ell}
  \left(b_{\mathrm{s}}r_{\mathrm{s}+}
  -b_{\mathrm{b}}r_{\mathrm{b}+}\right) 
  \end{aligned}
  \label{eq:regulated-free-energy}
\end{equation}
using \eqref{eq:periodicity_matching}.
  The first term is the standard soliton-black hole contribution, the second term is the cutoff-sensitive string cloud contribution, and the third term is the finite string cloud contribution. However, a subtle issue arises when the string cloud parameters of the black hole and soliton are allowed to differ. After imposing the matching conditions and taking the large-$R$ limit, the free energy density difference contains a term that grows linearly with the cutoff radius $R$. Unlike the overall volume divergence associated with the non-compact planar directions, this divergence is a genuine UV divergence and cannot be removed by a subtraction procedure.

The appearance of a cutoff-dependent contribution therefore suggests that two geometries do not share the same asymptotic string cloud background when $b_b \neq b_s$. To eliminate this UV divergence and obtain a well-defined thermodynamic ensemble, we impose the condition $b_b = b_s$. Under this identification, the divergent contribution vanishes, and the resulting free energy density difference turns into finite in the limit $R \to \infty$.
From a thermodynamic perspectice, the allows for a meaningful comparison between the black hole and soliton phases. 

When $b_{\mathrm{b}}=b_{\mathrm{s}}=b$, we thus obtain
\begin{equation}
  \Delta F
  =
  \frac{\beta_{\mathrm{s}}}{\ell^3}
  \left(r_{\mathrm{s}+}^{3}-r_{\mathrm{b}+}^{3}\right)
  +\frac{b\beta_{\mathrm{s}}}{\ell}
  \left(r_{\mathrm{s}+}-r_{\mathrm{b}+}\right).
  \label{eq:delta_F_b}
\end{equation}
 Thus the string cloud does not disappear from the finite thermodynamics; rather, only its divergent contribution vanished when the two saddles carry the same cloud strength.

\section{Phase Structure with Positive and Negative String Cloud Parameters}

Having obtained the finite free energy density difference between the planar black hole and AdS soliton phases, we now investigate the influence of the string cloud parameter $b$ on the phase structure of the system. In the absence of a string cloud, the phase transition is determined solely by the relative sizes of the black hole horizon and the soliton tip. The introduction of the string cloud modifies this competition and may significantly alter the thermodynamic behavior. Since positive and negative values of $b$ lead to qualitatively different effects, we discuss these two cases separately in the following subsections.

\subsection{Positive String Cloud Parameter}

We first consider the case of a   string cloud with $b>0$.
Substituting ~\eqref{eq:inverse_T} into ~\eqref{eq:delta_F_b} we obtain
\begin{equation}
\Delta F
=
\frac{4\pi r_{s+}}
{\ell\left(3r_{s+}^2-b\ell^2\right)}
\Big[
(r_{s+}^3-r_{b+}^3)
+b\ell^2(r_{s+}-r_{b+})
\Big].
\label{eq:delta_F_b+}
\end{equation}
Since the temperature $T$ must be positive, $\beta_s$ and $\beta_b
$ must also be positive. Therefore, from~\eqref{eq:inverse_T}, we obtain
\begin{equation}\label{eq:rs_rb_constrain}
r_{s+}>\frac{\sqrt{b}\ell}{\sqrt{3}},\qquad r_{b+}>\frac{\sqrt{b}\ell}{\sqrt{3}}.
\end{equation}
which implies that both the soliton tip radius and the black hole horizon radius must exceed this critical value. 

\begin{figure}[htbp]
    \centering
    \includegraphics[width=0.7\textwidth]{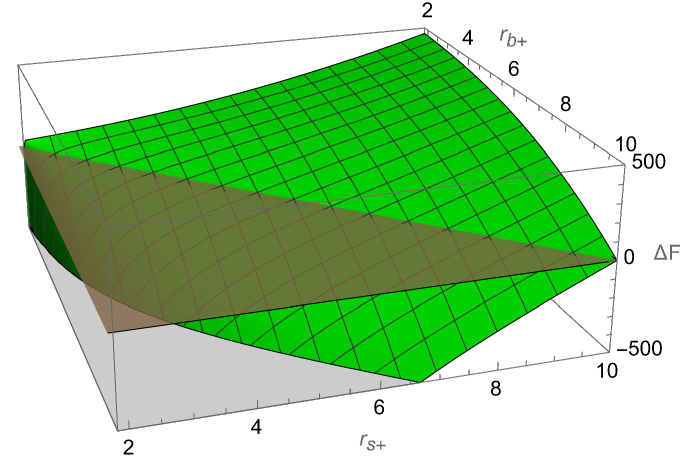}
    \caption{Free energy landscape for a positive string cloud parameter $b=10$. The free energy difference $\Delta F$ is plotted as a function of $r_{b+}$ and $r_{s+}$ with $l=1$. The shaded plane denotes $\Delta F=0$, and its intersection with the free energy surface marks the phase transition between the black hole and soliton phases.}
    \label{fig:2}
\end{figure}

To gain a clearer understanding of the dominant phase and the detailed structure of the phase transition, we plot the free energy difference as a function of the black hole horizon radius $r_{b+}$ and the soliton tip radius $r_{s+}$. For illustration, we fix the parameters $\ell=1$ and $b=10$. The resulting free energy landscape is shown in Fig.~\ref{fig:2}. From this figure, we can directly identity the regions in which the black hole or soliton phase is thermodynamically preferred, and the location of the sign of free energy difference changes.
As shown in Fig.~\ref{fig:2}, the free energy difference is poltted as the function of the black hole horizon radius $r_{b+}$ and soliton tip radius $r_{s+}$. Each point on the surface corresponds a particular pair of values $r_{b+}$, $r_{s+}$ and the associated free energy difference. 
It is also evident from~\eqref{eq:delta_F_b+} that {when $r_{b+} > r_{s+}$, the} free energy difference is negative, implying that the free energy of the black hole is lower than that of the soliton. Consequently, the black hole phase is thermodynamically perferred. {Conversely, when $r_{b+}<r_{s+}$, the} free energy difference becomes positive, indicating that the soliton has the lower free energy and therefore dominates the thermodynamic ensemble. These results are {commensurate with those obtained in the
$b=0$ case \cite{surya2001phase}: small  black holes in a string gas decay into solitons, whereas large black holes are thermodynamically stable.} 

\begin{figure}[htbp]
    \centering
    \includegraphics[width=0.7\textwidth]{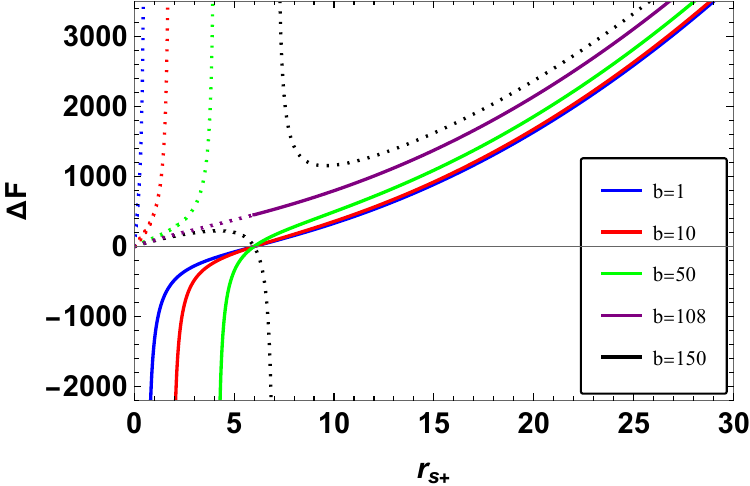}
    \caption{Free energy difference as a function of the soliton tip radius $r_{s+}$ for fixed $\ell=1$ and $r_{b+}=6$. The blue, red, green, purple, and black curves correspond to $b=1$, $10$, $50$, $108$, and $150$, respectively. The dashed segments indicate regions that violate the positivity condition of the temperature and are therefore excluded from the physical parameter space.
    }
    \label{fig:3}
\end{figure}

To investigate the effect of the string cloud parameter $b$ on the free energy difference, we fix $\ell=1$ and $r_{b+}=6$, and evaluate the free energy difference for several representative values of $b$. The results are shown in Fig.~\ref{fig:3}.

As shown in Fig.~\ref{fig:3}, the free energy difference exhibits a strong dependence on the string cloud parameter $b$. The dashed portions of the curves correspond to regions with negative periods, which violate the regularity condition and are therefore excluded from the physical parameter space. The horizontal line $\Delta F=0$ separates the parameter space into two thermodynamic phases. Regions with $\Delta F<0$ correspond to black hole phase, where the black hole possesses a lower free energy and is thermodynamically preferred. Conversly, regions with $\Delta F>0$ correspond to the soliton phase.

A particularly interesting case 
{occurs for $b\ell^2=3r_b^2$, when the black hole is extremal. 
The free energy becomes
\begin{equation} \label{eq:delta_F_extremal}
\Delta F
=
\frac{4\pi r_{s+}}{3}
\frac{4 r_{b+}^2 + r_{b+} r_{s+} + r_{s+}^2}{r_b+r_s}
\end{equation}
which is always greater than zero,  indicating that the soliton phase is always thermodynamically preferred over an extremal black hole.
This case is illustrated by the purple curve in Fig.~\ref{fig:3}, with $b=108$. 
We can understand this by writing $x=r_{b+}/r_{s+}$ and $b\ell^2 = \textsf{b}r^2_{b+}$, in which case 
\begin{equation} \label{eq:delta_F_extremal_b}
\Delta F
=
 \frac{4\pi r^2_{s+}}{\ell}
\frac{(x-1)((\textsf{b}+1)x^2+x+1)}{\textsf{b}x^2-3}
\end{equation}
where    $0<x<\sqrt{3/\textsf{b}}$ due to the constraint \eqref{eq:rs_rb_constrain}. The function on the right hand side of \eqref{eq:delta_F_extremal_b} is positive
for $0\leq x <1$, where the soliton phase is stable, and is negative for $1<x<\sqrt{3/\textsf{b}}$, where the black hole phase is stable.
As $\textsf{b}\to 3$ this second region diminishes in size, vanishing for $\textsf{b}=3$. In this case the black hole is extremal and the  soliton is the only stable phase.
If $\textsf{b} > 3$ the entire solution lies in an unphysical region of negative temperature; this is illustrated by the $b=150$ black curve in
Fig.~\ref{fig:3}.
}

\subsection{Negative String Cloud Parameter}
We now turn to the case of a negative string cloud parameter. From \eqref{eq:stringdensity}, the string cloud density is proportional to $b$, implying that negative values of $b$ correspond to a negative energy density. Consequently, the associated stress tensor violates the usual energy conditions. Nevertheless, it is still instructive to investigate this regime, since reversing the sign of the string cloud contribution may significantly modify the thermodynamic properties of the system and lead to qualitatively different phase structures.

We now consider the case of a negative string cloud parameter. In contrast to the positive $b$ case, the regularity condition does not impose a lower bound on the horizon radius or the soliton tip radius. Setting $b=-|b|$, the corresponding period remains positive for all positive values of $r_{b+}$ and $r_{s+}$,
\begin{equation}
\begin{aligned}
\beta_{b}
=\frac{4\pi r_{b+}\ell^2}{3r_{b+}^{2}+|b|\ell^{2}}>0 ,\\
\beta_{s}
=\frac{4\pi r_{s+}\ell^2}{3r_{s+}^{2}+|b|\ell^{2}}>0 ,
\end{aligned}
\end{equation}
and no extremal configuration arises. Consequently, the entire parameter space is thermodynamically admissible, allowing us to explore the influence of a negative string cloud parameter on the phase structure without the restrictions encountered in the positive-$b$ regime.

Accordingly, the free energy difference becomes
\begin{equation}\label{eq:delta_F_b-}
\Delta F
=
\frac{4\pi r_{s+}}
{\ell\left(3r_{s+}^2+|b|\ell^2\right)}
\Big[
(r_{s+}^3-r_{b+}^3)
-|b|\ell^2(r_{s+}-r_{b+})
\Big].
\end{equation}
A qualitatively new feature emerges in the negative-$b$ regime. Unlike the positive string cloud case, the free energy difference is no longer determined solely by the relative sizes of the black hole horizon and the soliton tip. Two competing contributions appear in the expression for $\Delta F$. The first is proportional to $r_{s+}-r_{b+}$, reproducing the behavior found for positive $b$. The second contribution, proportional to $|b|\ell^{2}$, enters with the opposite sign and competes with the geometric term. As a result, the sign of the free energy difference is no longer controlled exclusively by the relative sizes of $r_{s+}$ and $r_{b+}$. Consequently, the dominant thermodynamic phase may be altered by the presence of a negative string cloud parameter, leading to a phase structure that differs significantly from the positive-$b$ case.

\begin{figure}[htbp]
    \centering
    \includegraphics[width=0.7\textwidth]{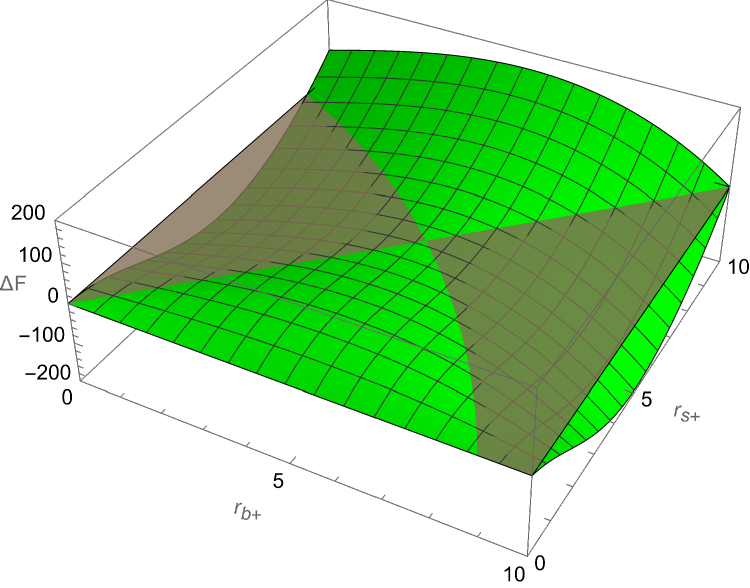}
    \caption{Free energy landscape for a positive string cloud parameter $b=-80$. The free energy difference $\Delta F$ is plotted as a function of $r_{b+}$ and $r_{s+}$ with $l=1$. The shaded plane denotes $\Delta F=0$, and its intersection with the free energy surface marks the phase transition between the black hole and soliton phases.}
    \label{fig:4}
\end{figure}

The corresponding free energy landscape for a negative string cloud parameter is shown in Fig.~\ref{fig:4}. Compared with the positive-$b$ case, the structure of the free energy surface changes dramatically. For positive values of the string cloud parameter, the sign of the free energy difference is determined entirely by the factor $r_{s+}-r_{b+}$, and the phase transition therefore occurs at $r_{s+}=r_{b+}$. In contrast, when $b<0$, the free energy difference contains an additional competing contribution. Setting $\Delta F=0$ yields
\begin{equation}
r_{s+}=r_{b+}\quad \textrm{and} \quad
r_{s+}=\frac{-r_{b+}+\sqrt{4|b|\ell^{2}-3r_{b+}^{2}}}{2}.
\end{equation}
Therefore, the phase boundary is no longer described by a single critical curve. The emergence of this additional branch explains the multiple intersections between the free energy surface and the plane $\Delta  F=0$ shown in Fig.~\ref{fig:4}.  The negative string cloud contribution introduces an additional competing scale, resulting in a significantly richer phase structure than that found in the positive-$b$ regime. 

\begin{figure}[htbp]
    \centering
    \includegraphics[width=0.7\textwidth]{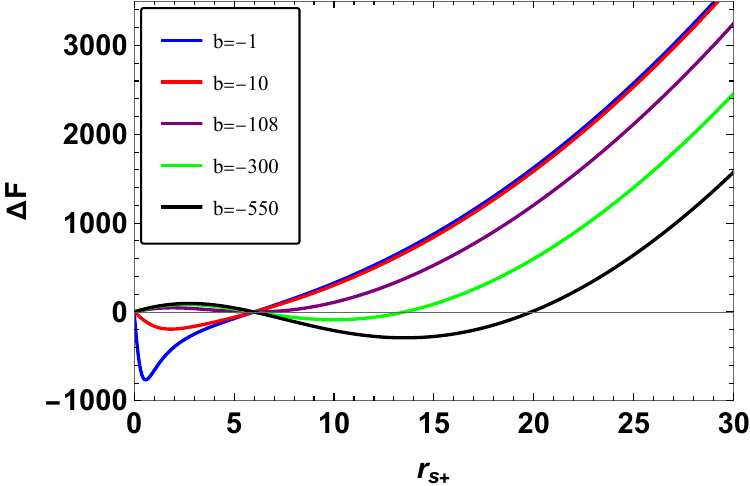}
    \caption{Free energy difference as a function of the soliton tip radius $r_{s+}$ for fixed $\ell=1$ and $r_{b+}=6$. The blue, red, purple, green, and black curves correspond to $b=-1$, $-10$, $-108$, $-300$, and $-550$, respectively.}
    \label{fig:5}
\end{figure}

To further illustrate the effect of the negative string cloud parameter, we fix $\ell=1$ and $r_{b+}=6$, and plot the free energy difference for several values of $b$, as shown in Fig.~\ref{fig:5}. The purple curve corresponds to the critical value $b=b_c$, which can be obtained by,
\begin{equation}
|b_c|
=
\frac{3r_{b+}^2}{\ell^2}.
\end{equation}
For $|b|<|b_c|$, the phase structure is similar to the positive-$b$ case: the dominant phase is mainly determined by the relative sizes of $r_{b+}$ and $r_{s+}$. The most interesting behavior appears when $|b|>|b_c|$, where the additional critical branch becomes physical. In this regime, the free energy difference vanishes at two distinct values of $r_{s+}$.

{We observe for  a fixed value of $b<0$ and $r_{b+} <  \frac{-r_{s+}+\sqrt{4|b|\ell^{2}-3r_{s+}^{2}}}{2}$ that  $\Delta  F>0$ for   small $r_{s+}$.  
The
 stable phase of the system is thus a soliton, and
 moves from soliton to black hole to soliton as $r_{s+}$ monotonically increases. Alternatively, for large $r_{b+}$  and 
 fixed $r_{s+} > \frac{-r_{b+}+\sqrt{4|b|\ell^{2}-3r_{b+}^{2}}}{2}$,  the stable phase of the system moves from black hole to soliton to black hole as $r_{b+}$ monotonically decreases.} 

Therefore, as $r_{b+}$ decreases, the system undergoes the sequence
\[
\text{Black Hole}
\rightarrow
\text{Soliton}
\rightarrow
\text{Black Hole}.
\]
This behavior is qualitatively different from that observed in the positive-$b$ regime, where only a single phase transition occurs. The reappearance of the black hole phase at large $r_{s+}$ is an example of a reentrant  phase transition \cite{altamirano2013reentrant}, and originates from the competition between the geometric contribution and the negative string cloud contribution in the free energy difference.

The critical case $b=b_c$ is particularly important. In this case, the free energy difference touches the $\Delta F=0$ line but does not develop a region where the black hole phase dominates; {in Fig.~\ref{fig:4}, this occurs along the line $r_{b+} = 5.164$.}
 Therefore, the soliton phase remains thermodynamically preferred throughout the physical parameter range. 
 {Alternatively, the black hole  remains thermodynamically preferred throughout the physical parameter range if $r_{s+} = 5.164$ in Fig.~\ref{fig:4}.}
 This critical value marks the boundary between the ordinary phase structure and the regime in which the negative string cloud parameter generates an additional transition branch.

\section{Conclusion}

We have studied the phase transition between planar AdS black holes and AdS solitons in the presence of a string cloud. The string cloud was modeled as a continuous distribution of radial fundamental strings, which forms a brush-like configuration in the planar background. The backreaction of the string cloud modifies the bulk geometry and introduces a new parameter into the thermodynamic competition between the black hole and soliton phases.

By comparing the Euclidean on-shell actions of the two geometries under the same asymptotic boundary conditions, we derived the corresponding free energy difference. We showed that if the string cloud parameters carried by the black hole and soliton are allowed to differ, the free energy difference contains a cutoff-dependent term that diverges linearly with the large-$R$ cutoff. Therefore a well-defined ensemble is recovered only when the string cloud parameters are identified, in which case the divergent contribution cancels and the free energy difference remains finite in the large-$R$ cutoff limit.

For positive values of the string cloud parameter, the sign of the free energy difference is determined by the relative sizes of the black hole horizon radius and the soliton tip radius, as in the $b=0$ case \cite{surya2001phase}.  The transition occurs at $r_{b+}=r_{s+}$, separating the black hole-dominated and soliton-dominated regions. The positivity of the temperature further imposes a lower bound on both radii, leading to the appearance of extremal configurations and restricting the physical parameter space. {The qualitative structure of the phase transition remains similar to that of the ordinary soliton-black hole transition,  but with one key difference: the soliton phase is thermodynamically preferred over an extremal black hole. }

The situation changes significantly when the string cloud parameter becomes negative. In this case, the temperature remains positive throughout the parameter space and no extremal bound appears. More importantly, the free energy difference develops an additional critical branch besides the conventional transition line $r_{b+}=r_{s+}$. As a consequence, the dominant phase is no longer determined solely by the relative sizes of the black hole horizon and soliton tip. Instead, the negative string cloud contribution introduces an additional competing scale that qualitatively modifies the phase structure.

For sufficiently large negative values of the string cloud parameter, the system exhibits the reentrant sequence
\[
\text{Black Hole}
\rightarrow
\text{Soliton}
\rightarrow
\text{Black Hole},
\]
as the soliton tip radius increases.  We also identified a critical value of the string cloud parameter that separates the conventional phase structure from the regime where the additional transition branch becomes physical.

Our work demonstrates that the presence of a string cloud  plays a crucial role in determining the thermodynamic behaviour of the system. While positive energy string clouds mainly shift the location of the transition, negative energy string clouds can qualitatively change the phase structure and generate new thermodynamic phenomena. Similar effects can be expected to persist in higher-dimensional spacetimes.
It would be interesting to investigate how higher-curvature corrections and the presence of electromagnetic charge further modifies the thermodynamic behaviour of this system.

\acknowledgments

This work was supported in part by the Natural Sciences and Engineering Research Council of Canada.




\bibliographystyle{JHEP}
\bibliography{biblio}

\end{document}